\documentclass[11pt]{article}
\usepackage{amssymb}
\usepackage{amsmath}
\usepackage[color,matrix,arrow,all]{xy}
\usepackage{times}
\usepackage{graphicx}
\usepackage{xcolor}

\title{Wadge degrees of $\bortwo$ omega-powers\\}
\author{to appear in Mathematical Logic Quarterly\\ \\ \\ 
Olivier FINKEL$^1$  and Dominique LECOMTE$^2$}
\date{\today}

\def\ufootnote#1{\let\savedthfn\thefootnote\let\thefootnote\relax
\footnote{#1}\let\thefootnote\savedthfn\addtocounter{footnote}{-1}}

\newcommand{\ol}{ $\omega$-language}

\newcommand{\om}{\omega}

\newcommand{\boraone}{{\bf\Sigma}^{0}_{1}}

\newcommand{\boraxi}{{\bf\Sigma}^{0}_{\xi}}

\newcommand{\borone}{{\bf\Delta}^{0}_{1}}
\newcommand{\bortwo}{{\bf\Delta}^{0}_{2}}

\newcommand{\bormone}{{\bf\Pi}^{0}_{1}}

\newcommand{\bormxi}{{\bf\Pi}^{0}_{\xi}}

\newcommand{\borxi}{{\bf\Delta}^{0}_{\xi}}
\newcommand{\borme}{{\bf\Pi}^{0}_{\eta}}

\newtheorem{thm} {Theorem} [section]
\newtheorem{defi} [thm] {Definition}
\newtheorem{rem} [thm] {Remark}

\newtheorem{lem} [thm] {Lemma}

\begin{document}

\maketitle

\centerline{$\bullet^1$  CNRS,  Universit\'e Paris Cit\'e and Sorbonne Universit\' e, IMJ-PRG, F-75013 Paris, France}

\centerline{finkel@math.univ-paris-diderot.fr}\medskip

\centerline{$\bullet^2$ 1) Sorbonne Universit\' e  and  Universit\'e Paris Cit\'e, CNRS, IMJ-PRG, F-75005 Paris, France}

\centerline{dominique.lecomte@upmc.fr}\medskip

\centerline{$\bullet^2$ 2) Universit\'e de Picardie, I.U.T. de l'Oise, site de Creil,}

\centerline{13, all\'ee de la fa\"\i encerie, 60100 Creil, France}

\medskip\medskip\medskip\medskip\medskip\medskip

\ufootnote{{\it 2020 Mathematics Subject Classification.}~Primary: 03E15, 68Q45; Secondary: 54H05, 03D05}

\ufootnote{{\it Keywords and phrases.}~complete set, difference of sets, Lavrentieff hierarchy, omega-power, 
omega-regular language, regular language, Wadge hierarchy, Wagner hierarchy}

\noindent {\bf Abstract.} We provide, for each natural number $n$ and each class among $D_n(\boraone )$, 
$\check D_n(\boraone )$ and $D_{2n+1}(\boraone )\oplus\check D_{2n+1}(\boraone )$, a regular language whose associated omega-power is complete for this class.

\vfill\eject

\section{$\!\!\!\!\!\!$ Introduction}\indent

 In the sixties, in order to prove the decidability of the monadic second order theory of one successor over the integers, B\"uchi  studied acceptance of infinite words by finite automata with the now called  B\"uchi  acceptance condition, see \cite{Buchi62}. Since then, a lot of work has been done on regular $\omega$-languages, accepted by B\"uchi automata, or by some other variants of automata over infinite words, like Muller or Rabin automata, see \cite{Thomas90,Staiger97,PerrinPin}.\medskip

 The class of regular $\om$-languages, those accepted by B\"uchi or Muller automata, is the $\om$-Kleene closure of the class of regular finitary languages. Let $\Sigma$ be a finite alphabet, and $L$ be a finitary language over 
$\Sigma$. The $\om$-{\bf power} $L^\infty$ of $L$ is the set of infinite words constructible with $L$ by concatenation, i.e., 
$L^\infty\! :=\!\{\ w_0w_1\ldots\!\in\!\Sigma^\omega\mid\forall i\!\in\!\omega~\ w_i\!\in\! L\ \}$. 
The $\om$-{\bf Kleene closure} of a class $\mathcal{C}$ of languages of finite words over finite alphabets is the class of $\om$-languages of the form 
${\bigcup_{1\leq j \leq n}~K_j\cdot L_j^\infty}$, where $n$ is a natural number and the 
$K_j$'s  and the $L_j$'s are in $\mathcal{C}$. We denote here by $L^\infty$ the 
$\omega$-power of $L$, as in \cite{Lecomte-JSL,Fink-Lec2} and the recent survey paper  \cite{FinkelLecomte20chapter}, while it is usually denoted by $L^\om$ in theoretical computer science papers, as in \cite{Staiger97,Fin01a,Fin03a,Fin-Lec}.\medskip
      
 Moreover, the operation of taking the $\om$-power of a finitary language also appears in the characterization of the class of context-free\ol s as the $\om$-Kleene closure of the family  of context-free finitary languages, see \cite{Staiger97}. \medskip

 This shows that the $\om$-power operation is a fundamental operation over finitary languages in the study of \ol s, which naturally leads to the question of its complexity. Since the set $\Sigma^\omega$ of infinite words over the finite alphabet $\Sigma$ can be  equipped with the usual Cantor topology, the question of the topological complexity of $\om$-powers of finitary languages naturally arises, and was asked by Niwinski \cite{Niwinski90},  Simonnet \cite{Simonnet92}, and Staiger \cite{Staiger97}.\medskip

 Then the  $\om$-powers have  been studied from the perspective of Descriptive Set Theory in several recent papers \cite{Fin01a,Fin03a,Fin04-FI,Lecomte-JSL,Fin-Dup06,Fin-Lec,Fink-Lec2,FinkelLecomte21,FinkelLecomte20chapter}.  \medskip

 As noticed by Simonnet in \cite{Simonnet92}, the $\om$-powers are always analytic sets. It has been proved in \cite{Fin03a} that there exists a finitary language $L$ accepted by a one counter automaton such that the $\om$-power $L^\infty$ is analytic and non Borel. Moreover, we proved in \cite{Fink-Lec2} that, for every non-null countable ordinal $\xi$, we can find ${\bf\Sigma}^{0}_{\xi}$-complete $\om$-powers, as well as 
 ${\bf\Pi}^{0}_{\xi}$-complete $\om$-powers. This shows that, surprisingly, the $\om$-powers can be very complex. Moreover, some results have been obtained in 
\cite{Fink-Lec2} about the Wadge degrees of $\om$-powers, indicating the location of some  $\om$-powers inside the Wadge hierarchy, which is a great refinement of the Borel hierarchy, see \cite{Wadge83}.\medskip

On the other hand,  there were very few results about the Wadge degrees  of  $\om$-powers of very low Borel rank.   We fill this gap in this paper, studying the  $\om$-powers in the class $\bortwo$.\medskip

 Our main result is the following:\medskip

\noindent\bf Theorem.\it \label{infclas} Let $n$ be a natural number.\smallskip

(a) We can find a regular language $L\!\subseteq\! 2^{<\omega}$ such that $L^\infty$ is complete for the class $D_n(\boraone )$, and another one for $\check D_n(\boraone )$.\smallskip

(b) We can find a regular language $L\!\subseteq\! 2^{<\omega}$ with the property that 
$L^\infty$ is complete for the class $D_0(\boraone )\oplus\check D_0(\boraone )$, and another one for $D_{2n+1}(\boraone )\oplus\check D_{2n+1}(\boraone )$.\rm\medskip

 Up to our knowledge, the $\omega$-powers we get are the first examples having these topological complexities, even in the non-necessarily regular case (except, for (a) when $n\!\leq\! 2$ and $\check D_3(\boraone )$, and for (b) when $n\! =\! 0$, see \cite{Lecomte-JSL}).\medskip

 Moreover, our $\om$-powers are $\om$-powers of regular finitary languages. The $\om$-powers of this kind have been studied by Litovsky and Timmerman in \cite{LitovskyT87}, where they proved that if a regular $\om$-language $\mathcal{L}$  is an $\om$-power, then there is a regular finitary language $L$ such that $\mathcal{L}\! =\! L^\infty$.\medskip 

The topological complexity of regular $\om$-languages is  well known. Every regular 
$\om$-language is a finite boolean combination of ${\bf \Pi}^0_2$ (and hence Borel) sets, and is in particular a ${\bf \Delta}_3^0$ set. The trace of the Wadge hierarchy on the 
$\omega$-regular languages is called the Wagner hierarchy. It has been completely described by Wagner in \cite{Wagner79}, see also \cite{Staiger97,Selivanov98,Selivanov08m}. Its length is the (countable) ordinal 
$\omega^\omega$.\medskip 

In particular, a regular $\om$-language is in the class $\bortwo$ iff it is in the hierarchy of finite differences of  $\boraone$ sets. Our main result implies that, for any Wagner class 
${\bf\Gamma}\!\not=\!\check {\bf\Gamma}$ with 
${\bf\Gamma}\!\subseteq\!\bortwo$, we can find $L\!\subseteq\! 2^{<\omega}$ regular such that $L^\infty$ is complete for $\bf\Gamma$. As a consequence, we determine the Wadge-Wagner hierarchy of non self-dual $\bortwo$ regular $\om$-powers.   

\section{$\!\!\!\!\!\!$ Preliminaries}\indent

We assume that the reader has a knowledge of basic notions of descriptive set theory, which can be found in \cite{Kechris94}.  We  use in this paper  usual notations in this field. 
  In the sequel, $\bf\Gamma$, $\bf\Lambda$ will be classes of subsets of Polish spaces. The class of complements of elements of $\bf\Gamma$ is defined, when $X$ is a Polish space, by $\check {\bf\Gamma}(X)\! :=\!\{ X\!\setminus\! A\mid A\!\in\! {\bf\Gamma}(X)\}$.\medskip

 We set $\boraone\! :=\!\{ O\mid O\mbox{ is an open subset of a Polish space }X\}$. If $\xi\!\geq\! 1$ is a countable ordinal, then $\bormxi\! :=\!\check\boraxi$ and $\borxi\! :=\!\boraxi\cap\bormxi$. If $\xi\!\geq\! 2$ and $X$ is a Polish space, then, inductively, 
$$\boraxi (X)\! :=\!
\big\{\bigcup_{n\in\omega}~A_n\mid\forall n\!\in\!\omega ~~A_n\!\in\!\bigcup_{\eta <\xi}~\borme (X)\big\} .$$
The classes $\boraxi$ and $\bormxi$ form the {\bf Borel hierarchy}. This hierarchy can be refined, using the classes of differences. If $\zeta$ is a countable ordinal and $O\! :=\! (O_\eta )_{\eta <\zeta}$ is an increasing sequence of subsets of a set $X$, then 
$D_\zeta (O)\! :=\!\{ x\!\in\! X\mid\exists\eta\! <\!\zeta ~~\mbox{parity}(\eta )\!\not=\!\mbox{parity}(\zeta )\wedge 
x\!\in\! O_\eta\!\setminus\! (\bigcup_{\theta <\eta}~O_\theta )\}$. If $X$ is a Polish space, then 
$D_\zeta ({\bf\Gamma})(X)\! :=\!\{ D_\zeta (O)\mid\forall\eta\! <\!\zeta ~~O_\eta\!\in\! {\bf\Gamma}(X)\}$. The classes 
$D_\zeta (\boraxi )$ and $\check D_\zeta (\boraxi )$ form the {\bf Lavrentieff hierarchy}. We set 
$$({\bf\Gamma}\!\oplus\! {\bf\Lambda})(X)\! :=\!\{ (A\cap C)\cup (B\!\setminus\! C)\mid C\!\in\!\borone (X)\wedge 
A\!\in\! {\bf\Gamma}(X)\wedge B\!\in\! {\bf\Lambda}(X)\} .$$
By \cite{Wagner79}, the {\bf Wagner hierarchy} of $\bortwo$ sets is made of the classes $D_n(\boraone )$, 
$\check D_n(\boraone )$, as well as $D_n(\boraone )\!\oplus\!\check D_n(\boraone )$, where $n\!\in\!\omega$, and starts as follows:
$$\begin{matrix}  
& D_0(\boraone )\! =\!\{\emptyset\} & & D_1(\boraone )\! =\!\boraone & & D_2(\boraone ) & \ldots\cr 
& & D_0(\boraone )\!\oplus\!\check D_0(\boraone )\! =\!\borone & & \boraone\!\oplus\!\bormone & & \cr
& \check D_0(\boraone ) & & \check D_1(\boraone )\! =\!\bormone & & \check D_2(\boraone ) & \ldots
\end{matrix}$$

 We will also consider the class defined, when $X$ is a Polish space, by
$$({\bf\Gamma}\sqcap {\bf\Lambda})(X)\! :=\!
\{ A\cap B\mid A\!\in\! {\bf\Gamma}(X)\wedge B\!\in\! {\bf\Lambda}(X)\} .$$
Recall that $A\!\in\! {\bf\Gamma}(2^\omega )$ is ${\bf\Gamma}$-{\bf complete} if, for any 
$B\!\in\! {\bf\Gamma}(2^\omega )$, we can find $f\! :\! 2^\omega\!\rightarrow\! 2^\omega$ continuous such that 
$B\! =\! f^{-1}(A)$. Intuitively, this means that $A$ is part of the most complex sets in $\bf\Gamma$. The {\bf Wadge class} associated with $A$ is $\{ f^{-1}(A)\mid f\mbox{ continuous}\}$, which by inclusion of classes defines the {\bf Wadge hierarchy} mentioned in the introduction, which refines the Lavrentieff hierarchy. In particular, the Lavrentieff classes have complete sets. By Theorem 22.10 in \cite{Kechris94} and its proof, if 
${\bf\Gamma}\!\not=\!\check {\bf\Gamma}$, then $A\!\subseteq\! 2^\omega$ is $\bf\Gamma$-complete exactly when 
$A\!\in\! {\bf\Gamma}\!\setminus\!\check {\bf\Gamma}$.\medskip

 We now turn to basic notions of automata theory and formal language theory  in theoretical computer science, see for instance \cite{Staiger97,PerrinPin}. Let $\Sigma$ be a finite alphabet. Then 
$\Sigma^{<\omega}$ is the set of finite words over $\Sigma$. If $w\! :=\! a_1\cdots a_l\!\in\!\Sigma^{<\omega}$, then 
$\vert w\vert\! =\! l$ is the {\bf length} of $w$.

\begin{defi} A (finite) {\bf automaton} is a 5-tuple $\mathcal{A}\! :=\! (Q,\Sigma ,\delta ,q_0, F)$, where $Q$ is a finite set of states, $\Sigma$ is a finite alphabet, $\delta\! :\! Q\!\times\!\Sigma\!\rightarrow\! 2^Q$ is a map, $q_0\!\in\! Q$ is the initial state, and $F\!\subseteq\! Q$ is the set of final states.\smallskip

 Let $w\! :=\! a_1\cdots a_l\!\in\!\Sigma^{<\omega}$ be a finite word over $\Sigma$. A {\bf run of }$\mathcal{A}$ on 
$\sigma$ is a sequence $r\! :=\! (q_i)_{1\leq i\leq l+1}$ of states with $q_1\! =\! q_0$ and, for each $1\!\leq\! i\!\leq\! l$, $q_{i+1}\!\in\!\delta (q_i,a_i)$. The {\bf language accepted} by $\mathcal{A}$ is $L(\mathcal{A})\! :=\!
\{ w\!\in\!\Sigma^{<\omega}\mid\exists r\!\in\! Q^{\vert w\vert +1}~~r\mbox{ is a run of }\mathcal{A}\mbox{ on }w\mbox{ with }q_{l+1}\!\in\! F\}$. A language $L\!\subseteq\!\Sigma^{<\omega}$ is {\bf regular} if 
$L\! =\! L(\mathcal{A})$ for some  automaton $\mathcal{A}$.\smallskip

\end{defi}

\begin{defi} A {\bf B\"uchi automaton} is a 5-tuple $\mathcal{A}\! :=\! (Q,\Sigma ,\delta ,q_0, F)$, where $Q$ is a finite set of states, $\Sigma$ is a finite alphabet, $\delta\! :\! Q\!\times\!\Sigma\!\rightarrow\! 2^Q$ is a map, $q_0\!\in\! Q$ is the initial state, and $F\!\subseteq\! Q$ is the set of final states.\smallskip

 Let $\sigma\! :=\! a_1a_2\cdots\!\in\!\Sigma^\omega$. A {\bf run of }$\mathcal{A}$ on $\sigma$ is a sequence 
$r\! :=\! (q_i)_{i\geq 1}$ of states with $q_1\! =\! q_0$ and, for each $i\!\geq\! 1$, $q_{i+1}\!\in\!\delta (q_i,a_i)$. We set 
$\mbox{In}(r)\! :=\!\{ q\!\in\! Q\mid\exists^\infty i\!\geq\! 1~~q_i\! =\! q\}$. The $\omega$-{\bf language accepted} by 
$\mathcal{A}$ is $\mathcal{L}(\mathcal{A})\! :=\!\{\sigma\!\in\!\Sigma^\omega\mid\exists r\!\in\! Q^\omega ~~r\mbox{ is a run of }\mathcal{A}\mbox{ on }\sigma\mbox{ with }\mbox{In}(r)\cap F\!\not=\!\emptyset\}$. An $\omega$-language 
$\mathcal{L}\!\subseteq\!\Sigma^\omega$ is $\omega$-{\bf regular} if $\mathcal{L}\! =\!\mathcal{L}(\mathcal{A})$ for some B\"uchi automaton $\mathcal{A}$.\end{defi}

 The usual {\bf concatenation} of two finite words $v$ and $w$ is denoted $v\!\cdot\! w$, sometimes just $vw$. This concatenation is extended to the concatenation of a finite word $w$ and an $\omega$-word $\sigma$. The infinite word $w\!\cdot\!\sigma$ is then the $\omega$-word such that $(w\!\cdot\!\sigma )(k)\! =\! w(k)$ if $k\!\leq\!\vert w\vert$, and $(w\!\cdot\!\sigma )(k)\! =\!\sigma (k\! -\!\vert w\vert )$ if $k\! >\!\vert w\vert$. The concatenation can be extended in an obvious way to infinite sequences of finite words. The concatenation of a set $L$ of finite words with a set 
$\mathcal{L}$ of infinite words is the set of infinite words 
$L\!\cdot\!\mathcal{L}\! :=\!\{ w\!\cdot\!\sigma\mid w\!\in\! L\mbox{ and }\sigma\!\in\!\mathcal{L}\}$. 
The prefix relation is denoted by $\subseteq$: a finite word $v$ is a {\bf prefix} of a finite word $w$ (respectively, an infinite word $\sigma$), denoted $v\!\subseteq\! w$, if and only if there exists a finite word $w'$ (respectively, an infinite word $\sigma'$), such that $w\! =\! v\!\cdot\! w'$. The $\omega$-{\bf Kleene closure} of the family of regular languages is the class of $\omega$-languages of the form $\bigcup_{1\leq j\leq n}~K_j\!\cdot\! L_j^\infty$, for some regular languages $K_j$ and $L_j$, $1\!\leq\! j\!\leq\! n$. As mentioned in the introduction, the class of $\omega$-regular languages is the $\omega$-Kleene closure of the family of regular languages (see \cite{PerrinPin}).

\vfill\eject

\begin{thm} \label{chareg} (B\"uchi) Let $\Sigma$ be a finite alphabet, and $\mathcal{L}\!\subseteq\!\Sigma^\omega$ be an $\omega$-language. The following are equivalent:\smallskip

(a) $\mathcal{L}$ is $\omega$-regular,\smallskip

(b) we can find $n\!\in\!\omega$ and regular languages $K_j$, $L_j\!\subseteq\!\Sigma^{<\omega}$, 
$1\!\leq\! j\!\leq\! n$, with the property that $\mathcal{L}\! =\!\bigcup_{1\leq j\leq n}~K_j\!\cdot\! L_j^\infty$.\end{thm}

\section{$\!\!\!\!\!\!$ The proof of the main result}\indent

 We first give an inductive construction of the difference hierarchy.

\begin{lem} \label{set} Let $\bf\Gamma$ be a class of subsets of Polish spaces closed under finite intersections and finite unions, and $k$ be a natural number.\smallskip

(a) $\check D_k({\bf\Gamma})\sqcap {\bf\Gamma}\! =\! D_{k+1}({\bf\Gamma})$.\smallskip

(b) $\check D_{2k}({\bf\Gamma})\sqcap\check {\bf\Gamma}\! =\!\check D_{2k+1}({\bf\Gamma})$.\smallskip

(c) $\check D_{2k}({\bf\Gamma})\sqcap\check D_2({\bf\Gamma})\! =\!\check D_{2k+2}({\bf\Gamma})$.\smallskip

(d) $\big( D_{2k+1}({\bf\Gamma})\oplus\check D_{2k+1}({\bf\Gamma})\big)\sqcap\check D_2({\bf\Gamma})\! =\! 
D_{2k+3}({\bf\Gamma})\oplus\check D_{2k+3}({\bf\Gamma})$.\end{lem}

\begin{rem}{\rm 
The anonymous reviewer of this paper indicated us that ``items (a),
(b), (c) of Lemma \ref{set} were known for quite some time. Namely, they are
particular cases of Proposition 8 in Section 5 of  \cite{Selivanov85} about the
difference hierarchy over arbitrary bounded distributive lattice.  Moreover,
the proof of Proposition 8 is essentially the same as even earlier proof for
the particular case of difference hierarchy over the c.e. sets in Section 3, Proposition 2 of \cite{Ershov68}".\medskip   

However,  in order to keep the paper as self contained as possible for the reader, and because these results are used later in the sequel, we have kept our proofs of  items (a),
(b), (c), and give them below. }
\end{rem}

\noindent\bf Proof.\rm\ Fix a Polish space $X$.\medskip 

\noindent (a) Let $U\! :=\! (U_\eta )_{\eta <k}$ be an increasing sequence of subsets of $X$ in $\bf\Gamma$, 
$A\! :=\! X\!\setminus\! D_k(U)$, and $V$ be a subset of $X$ in $\bf\Gamma$. We set, for $\eta\! <\! k$, 
$O_\eta\! :=\! U_\eta\cap V$, and $O_k\! :=\! V$. Then $(O_\eta )_{\eta <k+1}$ is an increasing sequence of subsets of $X$ in $\bf\Gamma$, and 
$$\begin{array}{ll}
x\!\in\! A\cap V\!\!\!\!
& \Leftrightarrow\Big( x\!\notin\!\bigcup_{\eta <k}~U_\eta\vee
\exists\eta\! <\! k~~\big(\mbox{parity}(\eta )\! =\!\mbox{parity}(k)\wedge 
x\!\in\! U_\eta\!\setminus\! (\bigcup_{\theta <\eta}~U_\theta)\big)\Big)\wedge x\!\in\! V\cr
& \Leftrightarrow x\!\in\! D_{k+1}(O)\mbox{,}
\end{array}$$
proving that $A\cap V\!\in\! D_{k+1}({\bf\Gamma})$.\medskip

 Conversely, assume that $D_{k+1}(O)\!\in\! D_{k+1}({\bf\Gamma})$. We set 
$V\! :=\! O_k$ and, for $\eta\! <\! k$, $U_\eta\! :=\! O_\eta$, which implies that $U$ is an increasing sequence of subsets of $X$ in $\bf\Gamma$, $D_k(U)\!\in\! D_k({\bf\Gamma})$, $V$ is in $\bf\Gamma$, and 
$D_{k+1}(O)\! =\!\big( X\!\setminus\! D_k(U)\big)\cap V$ by the previous computation.\medskip

\noindent (b) Let $U\! :=\! (U_\eta )_{\eta <2k}$ be an increasing sequence of subsets of $X$ in $\bf\Gamma$, 
$A\! :=\! X\!\setminus\! D_{2k}(U)$, $V$ be a subset of $X$ in $\bf\Gamma$, and $B\! :=\! X\!\setminus\! V$. We set 
$O_0\! :=\! V$ and, for $\eta\! <\! 2k$, $O_{\eta +1}\! :=\! V\cup U_\eta$, so that $(O_\eta )_{\eta <2k+1}$ is an increasing sequence of subsets of $X$ in $\bf\Gamma$, and
$$\begin{array}{ll}
x\!\in\! A\cap B\!\!\!\!
& \Leftrightarrow\big( x\!\notin\!\bigcup_{\eta <2k}~U_\eta\vee
\exists j\! <\! k~~x\!\in\! U_{2j}\!\setminus\! (\bigcup_{\theta <2j}~U_\theta)\big)\wedge 
x\!\notin\! V\cr
& \Leftrightarrow x\!\notin\! D_{2k+1}(O)\mbox{,}
\end{array}$$
proving that $A\cap B\!\in\!\check D_{2k+1}({\bf\Gamma})$.

\vfill\eject

 Conversely, assume that $D_{2k+1}(O)\!\in\! D_{2k+1}({\bf\Gamma})$. We set $B\! :=\! X\!\setminus\! O_0$ and, for 
$\eta\! <\! 2k$, $U_\eta\! :=\! O_{\eta +1}$, which implies that $U$ is an increasing sequence of subsets of $X$ in 
$\bf\Gamma$, $D_{2k}(U)\!\in\! D_{2k}({\bf\Gamma})$, $B$ is in $\check {\bf\Gamma}$, and 
$X\!\setminus\! D_{2k+1}(O)\! =\!\big( X\!\setminus\! D_{2k}(U)\big)\cap B$ by the previous computation.\medskip

\noindent (c) The key fact is that 
$(C_1\!\setminus\! C_0)\cup (D_1\!\setminus\! D_0)\! =\! (C_1\cup D_1)\!\setminus\! (C_0\cap D_0)$ if 
$C_1\!\subseteq\! D_0$ and $D_1\!\subseteq\! C_0$. We may assume that $k\! >\! 0$.


 Let $U\! :=\! (U_\eta )_{\eta <2k}$ be an increasing sequence of subsets of $X$ in 
$\bf\Gamma$, $A\! :=\! X\!\setminus\! D_{2k}(U)$, $V\! :=\! (V_\eta )_{\eta <2}$ be an increasing sequence of subsets of $X$ in $\bf\Gamma$, and 
$B\! :=\! X\!\setminus\! D_2(V)$. We set $O_0\! :=\! V_0\cap U_0$, 
$O_1\! :=\! V_1\cap U_1$, 
$O_{2j}\! :=\! (V_0\cap U_{2j})\cup U_{2j-2}\cup (V_1\cap U_{2j-1})$ and 
$$O_{2j+1}\! :=\! (V_1\cap U_{2j+1})\cup U_{2j-1}$$ 
if $0\! <\! j\! <\! k$, $O_{2k}\! :=\! V_0\cup U_{2k-2}\cup (V_1\cap U_{2k-1})$ and 
$O_{2k+1}\! :=\! V_1\cup U_{2k-1}$, so that $(O_\eta )_{\eta <2k+2}$ is an increasing sequence of subsets of $X$ in $\bf\Gamma$, and
$$\begin{array}{ll}
x\!\in\! A\!\cap\! B\!\!\!\!\!
& \Leftrightarrow\!\big( x\!\notin\! U_{2k-1}\vee
\exists j\! <\! k~~x\!\in\! U_{2j}\!\setminus\! (\bigcup_{\theta <2j}~U_\theta)\big)\wedge 
x\!\in\!\big( V_0\cup (X\!\setminus\! V_1)\big)\cr
& \Leftrightarrow\! x\!\in\! V_0\!\setminus\! U_{2k-1}\vee
\exists j\! <\! k~~x\!\in\! (V_0\cap U_{2j})\!\setminus\! (\bigcup_{\theta <2j}~U_\theta )\vee
x\!\notin\! V_1\cup U_{2k-1}~\vee\cr
& \hfill{\exists j\! <\! k~~x\!\in\! U_{2j}\!\setminus\! (V_1\cup\bigcup_{\theta <2j}~U_\theta )}\cr
& \Leftrightarrow\! x\!\in\! V_0\!\cap\! U_0\vee\exists j\!\in\! (0,k)~
x\!\in\!\big( (V_0\!\cap\! U_{2j})\!\setminus\! U_{2j-1}\big)\!\cup\!
\big( U_{2j-2}\!\setminus\! (V_1\!\cup\!\bigcup_{\theta <2j-2}~U_\theta )\big)\cr
& \hfill{\vee x\!\in\!\big( V_0\!\setminus\! U_{2k-1}\cup 
U_{2k-2}\!\setminus\! (V_1\cup\bigcup_{\theta <2k-2}~U_\theta )\big)\vee
x\!\notin\! V_1\cup U_{2k-1}}\cr
& \Leftrightarrow\! x\!\in\! V_0\!\cap\! U_0\vee\exists j\!\in\! (0,k)~~
x\!\in\!\big( (V_0\!\cap\! U_{2j})\cup U_{2j-2}\big)\!\setminus\!\big( (V_1\!\cap\! U_{2j-1})\!\cup\!
\bigcup_{\theta <2j-2}~U_\theta\big)\cr
& \hfill{\vee x\!\in\! (V_0\cup U_{2k-2})\!\setminus\!\big( (V_1\cap U_{2k-1})\cup 
\bigcup_{\theta <2k-2}~U_\theta\big)\vee x\!\notin\! V_1\cup U_{2k-1}}\cr
& \Leftrightarrow\! x\!\notin\! D_{2k+2}(O)\mbox{,}
\end{array}$$
proving that $A\cap B\!\in\!\check D_{2k+2}({\bf\Gamma})$.\medskip

 Conversely, assume that $D_{2k+2}(O)\!\in\! D_{2k+2}({\bf\Gamma})$. We set, for $\eta\! <\! 2k$, 
$U_\eta\! :=\! O_\eta$, which implies that $U$ is an increasing sequence of subsets of $X$ in 
$\bf\Gamma$ and $D_{2k}(U)\!\in\! D_{2k}({\bf\Gamma})$. We also set, for $\eta\! <\! 2$, 
$V_\eta\! :=\! O_{2k+\eta}$, which implies that $V$ is an increasing sequence of subsets of $X$ in 
$\bf\Gamma$ and $D_2(V)\!\in\! D_2({\bf\Gamma})$. Moreover, 
$X\!\setminus\! D_{2k+2}(O)\! =\!\big( X\!\setminus\! D_{2k}(U)\big)\cap\big( X\!\setminus\! D_2(V)\big)$ by the previous computation.\medskip

\noindent (d) We first check that $({\bf\Gamma}\oplus\check {\bf\Gamma})\sqcap\check D_{2k}({\bf\Gamma})\! =\! 
D_{2k+1}({\bf\Gamma})\oplus\check D_{2k+1}({\bf\Gamma})$. Let $C$ be a clopen subset of $X$, $A$ be a subset of $X$ in ${\bf\Gamma}$, $B$ be a subset of $X$ in $\check {\bf\Gamma}$, and $E$ be a subset of $X$ in 
$\check D_{2k}({\bf\Gamma})$. Then 
$\big( (A\cap C)\cup (B\!\setminus\! C)\big)\cap E\! =\!\big( (E\cap A)\cap C\big)\cup\big( (E\cap B)\!\setminus\! C\big)$, showing one inclusion by (a) and (b). Conversely, let $D$ be a subset of $X$ in $D_{2k+1}({\bf\Gamma})$, and 
$F$ be a subset of $X$ in $\check D_{2k+1}({\bf\Gamma})$. By (a), we can find a subset $E_0$ of $X$ in 
$\check D_{2k}({\bf\Gamma})$ and a subset $A$ of $X$ in ${\bf\Gamma}$ with $D\! =\! E_0\cap A$. By (b), we can find a subset $E_1$ of $X$ in $\check D_{2k}({\bf\Gamma})$ and a subset $B$ of $X$ in $\check {\bf\Gamma}$ with $F\! =\! E_1\cap B$. Note that $(D\cap C)\cup (F\!\setminus\! C)\! =\!
\big( (E_0\cap A)\cap C\big)\cup\big( (E_1\cap B)\!\setminus\! C\big)\! =\!\big( (A\cap C)\cup (B\!\setminus\! C)\big)\cap\big( (E_0\cap C)\cup (E_1\!\setminus\! C)\big)$, showing the other inclusion. Indeed, 
$(E_0\cap C)\cup (E_1\!\setminus\! C)\!\in\!\check D_{2k}({\bf\Gamma})$ is true if $k\!\in\! 2$, and for $k\! >\! 2$ by induction, using (c) and this formula again.\medskip

 From this and (c) we deduce that 
$$\begin{array}{ll}
\big( D_{2k+1}({\bf\Gamma})\oplus\check D_{2k+1}({\bf\Gamma})\big)\sqcap\check D_2({\bf\Gamma})\!\!\!\!\!
& =\! ({\bf\Gamma}\oplus\check {\bf\Gamma})\sqcap\check D_{2k}({\bf\Gamma})\sqcap\check D_2({\bf\Gamma})\cr
& =\! ({\bf\Gamma}\oplus\check {\bf\Gamma})\sqcap\check D_{2k+2}({\bf\Gamma})\! =\!  
D_{2k+3}({\bf\Gamma})\oplus\check D_{2k+3}({\bf\Gamma})\mbox{,}
\end{array}$$ 
finishing the proof.\hfill{$\square$}\medskip
 
\noindent\bf Examples.\rm\ Here are three fundamental examples.\medskip

\noindent - If 
$L\! :=\!\{ w\!\in\! 2^{<\omega}\mid 0\!\subseteq\! w\vee\exists p\!\in\!\omega ~~10^p1\!\subseteq\! w\}$, then  
$L^\infty\! =\! 2^\omega\!\setminus\!\{ 10^\infty\}$ is $D_1(\boraone )$-complete.\medskip

\noindent - If $L\! :=\!\{ 0\}$, then $L^\infty\! =\!\{ 0^\infty\}$ is $\check D_1(\boraone )$-complete.\medskip

\noindent - If 
$L\! :=\!\{ w\!\in\! 2^{<\omega}\mid w\!\subseteq\! 0^\infty\vee\exists p,q\!\in\!\omega ~~0^p10^q1\!\subseteq\! w\}$,  then\medskip
 
\centerline{$L^\infty\! =\!\{\alpha\!\in\! 2^\omega\mid\alpha\! =\! 0^\infty\vee
\exists p\!\not=\! q~~\alpha (p)\! =\!\alpha (q)\! =\! 1\}$}\medskip
 
\noindent is $\check D_2(\boraone )$-complete. Indeed, it is enough to see that $L^\infty$ is not $D_2(\boraone )$. In order to see that, we argue by contradiction, which gives $O$ open and $C$ closed with $L^\infty\! =\! O\cap C$. Then $C$ must be $2^\omega$, and thus $L^\infty$ is open. The sequence $(0^n10^\infty )_{n\in\omega}$ gives the desired contradiction.\medskip

\noindent\bf Notation.\rm\ We set, for 
$\alpha\!\in\! 2^\omega$ and $\varepsilon\!\in\! 2$, 
$(\alpha )_\varepsilon\! :=\!(\alpha (\varepsilon ),\alpha (\varepsilon\! +\! 2),\alpha (\varepsilon\! +\! 4),\cdots )$. Similarly, if $w\!\in\! 2^{<\omega}$ has even length $2l$ and $\varepsilon\!\in\! 2$, then we set $(w)_\varepsilon\! :=\!\big( w(\varepsilon ),w(\varepsilon\! +\! 2),w(\varepsilon\! +\! 4),\cdots ,w(\varepsilon\! +\! 2l\! -\! 2)\big)$. We set, for 
$L\!\subseteq\! 2^{<\omega}$, $L^*\! :=\!\{ w_1\!\cdots\! w_l\mid l\!\in\!\omega\wedge\forall i\! <\! l~~w_{i+1}\!\in\! L\}$, and\medskip
 
\noindent - $L_0\! :=\!\big\{ w\!\in\! 2^{<\omega}\mid\vert w\vert\mbox{ is even }\wedge (w)_0\!\in\! L^*\wedge 
(0\!\subseteq\! (w)_1\vee\exists q\!\in\!\omega ~~10^q1\!\subseteq\! (w)_1)\big\}$,\medskip
 
\noindent - $L_1\! :=\!\big\{ w\!\in\! 2^{<\omega}\mid\vert w\vert\mbox{ is even }\wedge (w)_0\!\in\! L^*\wedge 
(w)_1\!\subseteq\! 0^\infty\big\}$,\medskip
 
\noindent - $L_2\! :=\!\big\{ w\!\in\! 2^{<\omega}\mid\vert w\vert\mbox{ is even }\wedge (w)_0\!\in\! L^*\wedge 
\big( (w)_1\!\subseteq\! 0^\infty\vee\exists p,q\!\in\!\omega ~~0^p10^q1\!\subseteq\! (w)_1\big)\big\}$. 

\begin{lem} \label{reg} Let $L\!\subseteq\! 2^{<\omega}$ be a regular language. Then $L_0$, $L_1$ and $L_2$ are also regular.\end{lem}

\noindent\bf Proof.\rm\    Recall first that the class of regular finitary languages over an alphabet $\Sigma\! =\!\{ a_1, a_2,\cdots , a_n\}$  is the closure of the class containing the emptyset and the singletons $\{a_i\}$ consisting of a single word of length $1$ (we identify here the letter $a_i$ with the word of length $1$ containing this single letter), under the operations of union, concatenation, and the star operation $L\!\mapsto\! L^*$ over finitary languages.  Then the class of finitary regular languages is also closed under intersection (and taking complements). These properties imply that if  
$L\!\subseteq\! 2^{<\omega}$ is a regular language, then $L_0$, $L_1$ and $L_2$ are also regular.\hfill{$\square$}\medskip

 The next lemma shows that the completeness can be propagated in the difference hierarchy.

\begin{lem} \label{propagation} Let $k$ be a natural number.\smallskip

(a) If there is $L\!\subseteq\! 2^{<\omega}$ such that $L^\infty$ is $\check D_k(\boraone )$-complete, then 
$L_0^\infty$ is $D_{k+1}(\boraone )$-complete.\smallskip

(b) If there is $L\!\subseteq\! 2^{<\omega}$ such that $L^\infty$ is $\check D_{2k}(\boraone )$-complete, then 
$L_1^\infty$ is $\check D_{2k+1}(\boraone )$-complete.\smallskip

(c) If there is $L\!\subseteq\! 2^{<\omega}$ such that $L^\infty$ is $\check D_{2k}(\boraone )$-complete, then 
$L_2^\infty$ is $\check D_{2k+2}(\boraone )$-complete.\smallskip

(d) If there is $L\!\subseteq\! 2^{<\omega}$ such that $L^\infty$ is complete for the class 
$D_{2k+1}({\bf\Sigma}^0_1)\oplus\check D_{2k+1}({\bf\Sigma}^0_1)$, then $L_2^\infty$ is complete for the class 
$D_{2k+3}(\boraone )\oplus\check D_{2k+3}(\boraone )$.\end{lem}

\noindent\bf Proof.\rm\ (a) Let us check that 
$L_0^\infty\! =\!\{\alpha\!\in\! 2^\omega\mid (\alpha )_0\!\in\! L^\infty\wedge (\alpha )_1\!\not=\! 10^\infty\}$. Assume that ${\alpha\!\in\! L_0^\infty}$, which gives a sequence 
$(w_i)_{i\in\omega}$ of nonempty words in $L_0$ with the property that 
$\alpha\! =\! w_0w_1\cdots$. Note that 
${(\alpha )_\varepsilon\! =\! (w_0)_\varepsilon (w_1)_\varepsilon\cdots}$ if 
$\varepsilon\!\in\! 2$ since the $\vert w_i\vert$'s are even. This implies that 
$(\alpha )_0\!\in\! L^\infty$ and $(\alpha )_1\!\not=\! 10^\infty$. Conversely, assume that these two properties hold. If $(\alpha )_1$ has finitely many $1$'s, then we choose an initial segment of $(\alpha )_0$ in $L^*$ starting a decomposition of $(\alpha )_0$ into words of $L$ of length $l$ large enough to ensure that $(\alpha )_1\vert l$ contains all the $1$'s in 
$(\alpha )_1$, and either $(\alpha )_1(0)\! =\! 0$, or we can find $p,q$ with 
$0^p10^q1\!\subseteq\! (\alpha )_1\vert l$.

\vfill\eject

 We set $(w_0)_\varepsilon\! :=\! (\alpha )_\varepsilon\vert l$ for each $\varepsilon\!\in\! 2$, so that $w_0\!\in\! L_0$. We then consider the rest of this decomposition of $(\alpha )_0$ into words of $L$, which gives $(w_{i+1})_0$. Setting 
$(w_{i+1})_1\! :=\! 0^{\vert (w_{i+1})_0\vert}$, we get $w_{i+1}\!\in\! L_0$ and 
$\alpha\! =\! w_0w_1\cdots$. If $(\alpha )_1$ has infinitely many $1$'s, then we construct $w_i\!\in\! L_0$, ensuring that $(w_i)_0\!\in\! L^*$ is long enough and 
$0^p10^q1\!\subseteq\! (w_i)_1$ for some $p,q\!\in\!\omega$. This proves that 
$\alpha\!\in\! L_0^\infty$. By Lemma \ref{set}.(a), $L_0^\infty\!\in\! D_{k+1}(\boraone )$. Assume now that $D$ is a $D_{k+1}(\boraone )$ subset of $2^\omega$. Lemma \ref{set}.(a) provides $C\!\in\!\check D_k(\boraone )$ and $O\!\in\!\boraone$ such that 
$D\! =\! C\cap O$. Let $f_0\! :\! 2^\omega\!\rightarrow\! 2^\omega$ be a continuous map with $C\! =\! f_0^{-1}(L^\infty )$, and $f_1\! :\! 2^\omega\!\rightarrow\! 2^\omega$ be a continuous map with $O\! =\! f_1^{-1}(2^\omega\!\setminus\!\{ 10^\infty\} )$. Then the map $f\! :\! 2^\omega\!\rightarrow\! 2^\omega$ defined by 
$\big( f(x)\big)_\varepsilon\! :=\! f_\varepsilon (x)$ is continuous and satisfies 
$D\! =\! f^{-1}(L_0^\infty )$, showing the completeness of $L_0^\infty$.\medskip

\noindent (b) Let us check that 
$L_1^\infty\! =\!\{\alpha\!\in\! 2^\omega\mid (\alpha )_0\!\in\! L^\infty\wedge (\alpha )_1\! =\! 0^\infty\}$. Assume that 
$\alpha\!\in\! L_1^\infty$, which gives a sequence $(w_i)_{i\in\omega}$ of nonempty words in $L_1$ with 
$\alpha\! =\! w_0w_1\cdots$. Then $(\alpha )_0\!\in\! L^\infty$ and $(\alpha )_1\! =\! 0^\infty$. Conversely, assume that these two properties hold. We set $(w_0)_0\! :=\! (\alpha )_0\vert l_0$, where $l_0\! >\! 0$ and $(w_0)_0\!\in\! L$ starts a decomposition of $(\alpha )_0$ into words of $L$. We set $(w_0)_1\! :=\! 0^{l_0}$, so that $w_0\!\in\! L_1$. We then continue this decomposition of $(\alpha )_0$ into words of $L$, which gives $(w_{i+1})_0$. We then set  
$(w_{i+1})_1\! :=\! 0^{\vert (w_{i+1})_0\vert}$, we get $w_{i+1}\!\in\! L_1$ and $\alpha\! =\! w_0w_1\cdots$, proving that $\alpha\!\in\! L_1^\infty$. By Lemma \ref{set}.(b), $L_1^\infty\!\in\!\check D_{2k+1}(\boraone )$. Assume now that $C$ is a $\check D_{2k+1}(\boraone )$ subset of $2^\omega$. Lemma \ref{set}.(b) provides 
$M\!\in\!\check D_{2k}(\boraone )$ and $P\!\in\!\bormone$ such that $C\! =\! M\cap P$. Let 
$f_0\! :\! 2^\omega\!\rightarrow\! 2^\omega$ be a continuous map with the property that $M\! =\! f_0^{-1}(L^\infty )$, and $f_1\! :\! 2^\omega\!\rightarrow\! 2^\omega$ be a continuous map with $P\! =\! f_1^{-1}(\{ 0^\infty\} )$. Then the map $f\! :\! 2^\omega\!\rightarrow\! 2^\omega$ defined by $\big( f(x)\big)_\varepsilon\! :=\! f_\varepsilon (x)$ is continuous and satisfies $C\! =\! f^{-1}(L_1^\infty )$, showing the completeness of $L_1^\infty$.\medskip

\noindent (c) Let us check that $L_2^\infty\! =\!\big\{\alpha\!\in\! 2^\omega\mid (\alpha )_0\!\in\! L^\infty\wedge\big( (\alpha )_1\! =\! 0^\infty\vee\exists p\!\not=\! q~~(\alpha )_1(p)\! =\! (\alpha )_1(q)\! =\! 1\big)\big\}$. Assume that 
$\alpha\!\in\! L_2^\infty$, which gives a sequence $(w_i)_{i\in\omega}$ of nonempty words in $L_2$ with 
$\alpha\! =\! w_0w_1\cdots$. As $(\alpha )_\varepsilon\! =\! (w_0)_\varepsilon (w_1)_\varepsilon\cdots$ if 
$\varepsilon\!\in\! 2$, $(\alpha )_0\!\in\! L^\infty$, and $\big( (\alpha )_1\! =\! 0^\infty\mbox{ or }
\exists p\!\not=\! q~~(\alpha )_1(p)\! =\! (\alpha )_1(q)\! =\! 1\big)$. Conversely, assume that these two properties hold. If $(\alpha )_1$ has finitely many $1$'s, then as in (a) we choose $l$ in such a way that $(\alpha )_1\vert l$ contains all the $1$'s in $(\alpha )_1$, and either $(w_0)_1\!\subseteq\! 0^\infty$, or $0^p10^q1$ is a prefix of $(w_0)_1$ for some $p,q\!\in\!\omega$. We conclude as in (a). If $(\alpha )_1$ has infinitely many $1$'s, then we argue as in (a) to see that $\alpha\!\in\! L_2^\infty$. By Lemma \ref{set}.(c), $L_2^\infty\!\in\!\check D_{2k+2}(\boraone )$. We set 
$$O_0\! :=\!\{\alpha\!\in\! 2^\omega\mid\exists p\!\not=\! q~~\alpha (p)\! =\!\alpha (q)\! =\! 1\}\mbox{,}$$ 
and $O_1\! :=\!\{\alpha\!\in\! 2^\omega\mid\alpha\!\not=\! 0^\infty\}$, so that $2^\omega\!\setminus\! D_2(O)$ is 
$\check D_2(\boraone )$-complete. Assume now that $C$ is a $\check D_{2k+2}(\boraone )$ subset of $2^\omega$. Lemma \ref{set}.(c) provides $M\!\in\!\check D_{2k}(\boraone )$ and $S\!\in\!\check D_2(\boraone )$ such that 
$C\! =\! M\cap S$. Let $f_0\! :\! 2^\omega\!\rightarrow\! 2^\omega$ be a continuous map with 
$M\! =\! f_0^{-1}(L^\infty )$, and $f_1\! :\! 2^\omega\!\rightarrow\! 2^\omega$ be a continuous map with 
$S\! =\! f_1^{-1}\big( 2^\omega\!\setminus\! D_2(O)\big)$. Then the map $f\! :\! 2^\omega\!\rightarrow\! 2^\omega$ defined by $\big( f(x)\big)_\varepsilon\! :=\! f_\varepsilon (x)$ is continuous and satisfies $C\! =\! f^{-1}(L_2^\infty )$, showing the completeness of $L_2^\infty$.\medskip

\noindent (d) The proof of (c) shows that 
$$L_2^\infty\! =\!\big\{\alpha\!\in\! 2^\omega\mid (\alpha )_0\!\in\! L^\infty\wedge\big( (\alpha )_1\! =\! 0^\infty\vee\exists p\!\not=\! q~~(\alpha )_1(p)\! =\! 
(\alpha )_1(q)\! =\! 1\big)\big\} .$$
By Lemma \ref{set}.(d), 
$L_2^\infty\!\in\! D_{2k+3}(\boraone )\oplus\check D_{2k+3}(\boraone )$. Assume now that 
$C$ is a subset of $2^\omega$ in 
$D_{2k+3}(\boraone )\oplus\check D_{2k+3}(\boraone )$. Lemma \ref{set}.(d) provides a subset $M$ of $2^\omega$ in 
$\big( D_{2k+1}(\boraone )\oplus\check D_{2k+1}(\boraone )\big)$, and 
$S\!\in\!\check D_2(\boraone )(2^\omega)$ such that $C\! =\! M\cap S$. We conclude as in (c).\hfill{$\square$}\medskip

\vfill\eject

\noindent\bf Proof of the main result.\rm\ (a) We argue by induction on $n$. For the class $D_0(\boraone )$, we can take $L\! :=\!\emptyset$, so that $L^\infty\! =\!\emptyset$. For 
$\check D_0(\boraone )$, we can take 
$L\! :=\! 2^{<\omega}$, so that $L^\infty\! =\! 2^\omega$. Then, using Lemma \ref{reg} and inductively, Lemma \ref{propagation}.(c) solves our problem for $\check D_{2k+2}(\boraone )$. Then Lemma \ref{propagation}.(a) solves our problem for $D_{2k+1}(\boraone )$, while Lemma \ref{propagation}.(b) solves our problem for 
$\check D_{2k+1}(\boraone )$. Now  Lemma \ref{propagation}.(a) solves our problem for $D_{2k+2}(\boraone )$.\medskip

\noindent (b) Note that $D_0(\boraone )\oplus\check D_0(\boraone )\! =\!\borone$. For 
$\borone$, we can take 
$L\! :=\!\{ w\!\in\! 2^{<\omega}\mid 0\!\subseteq\! w\vee 1^2\!\subseteq\! w\}$, so that 
$L^\infty\! =\! N_0\cup N_{1^2}$. Note then that 
$D_1(\boraone )\oplus\check D_1(\boraone )\! =\!\boraone\oplus\bormone$. For 
$\boraone\oplus\bormone$, we can take 
$L\! :=\!\{ 0^2,0^21\}\cup\{ w\!\in\! 2^{<\omega}\mid\exists p\!\in\!\omega ~~10^p1\!\subseteq\! w\}$, so that $L^\infty\! =\!
\{ 0^\infty\}\cap (\bigcup_{q\in\omega}~N_{0^{2q+2}1})\cup (N_1\!\setminus\!\{ 10^\infty\} )$. Then, inductively, Lemmas \ref{reg} and \ref{propagation}.(d) solve our problem for 
$D_{2n+3}(\boraone )\oplus\check D_{2n+3}(\boraone )$.\hfill{$\square$}

\section{$\!\!\!\!\!\!$ Concluding remarks}\indent

We proved that, for any  natural number $n$, we can find a  language 
$L\!\subseteq\! 2^{<\omega}$ such that $L^\infty$ is complete for the class $D_n(\boraone )$, and another one for $\check D_n(\boraone )$. On the other hand  the hierarchy of differences of open sets can be extended to transfinite ranks indexed by countable ordinals, and it is known that the class $\bortwo$ is actually the union of the classes $D_\xi(\boraone )$, where $\xi$ is a countable ordinal.  This naturally leads to the question of the existence of $\om$-powers located at an infinite level of the hierarchy of differences of open sets.\medskip

In the case of $\bortwo$ regular $\om$-powers, Wagner's study in \cite{Wagner79} shows  that they are all located inside the hierarchy of finite differences of open sets. So,  in order to determine which Wadge classes of $\bortwo$ sets $\bf\Gamma$ have the property that we can find $L\!\subseteq\! 2^{<\omega}$ regular such that $L^\infty$ is complete for $\bf\Gamma$, it remains to solve the question for the classes of the form $D_{2n+2}(\boraone )\oplus\check D_{2n+2}(\boraone )$, for $n\!\in\!\omega$. \bigskip 

{\bf \large Acknowledgments.}  We thank very much  the anonymous reviewer for very useful comments and for having indicated us the  references containing  previous proofs of Lemma \ref{set} (a), (b), (c).

\end{document}